\documentclass{moriond}

\def\Journal#1#2#3#4{{#1} {\bf #2}, #3 (#4)}

\def\NPB{{\em Nucl. Phys.} B}
\def\PLB{{\em Phys. Lett.}  B}

\def\PRD{{\em Phys. Rev.} D}

\def\JHEP{\em JHEP}

\def\be{\begin{equation}}
\def\ee{\end{equation}}
\def\bea{\begin{eqnarray}}
\def\eea{\end{eqnarray}}

\usepackage{ulem}
\usepackage{bbold}
\usepackage{float}
\usepackage{cancel}
\usepackage{subfig}
\usepackage{amsmath, amssymb, setspace,cite}
\usepackage{slashed}
\usepackage{graphicx}

\usepackage{soul}
\usepackage{xcolor}
\usepackage{braket}
\usepackage{makecell}
\usepackage{hhline}

\numberwithin{equation}{section}
\vspace{0.6cm}
\date{\today} 

\usepackage{tabularx}
\usepackage{longtable} 
\usepackage{multicol}
\usepackage{multirow}
\usepackage{enumitem}

\newcommand{\Photo}{}

\begin{document}
\vspace*{4cm}
\title{New LCSR predictions for hadronic form factors}

\author{Yann Monceaux}

\address{Universit\'e Claude Bernard Lyon 1, CNRS/IN2P3, \\
Institut de Physique des 2 Infinis de Lyon, UMR 5822, F-69622, Villeurbanne, France}

\maketitle
\begin{abstract}
We present an overview of a novel approach to the QCD Light-Cone Sum Rule method, employing $B$-meson Light-Cone Distribution Amplitudes. This method circumvents the semi-global quark-hadron duality (QHD) approximation, which can introduce an unknown and potentially significant systematic error in form factor predictions. Instead, our approach is more dependent on higher-order contributions in perturbation theory and higher-twist contributions. Unlike the systematic errors from semi-global QHD, truncation errors can be evaluated and systematically improved, allowing for robust form factor predictions.
\end{abstract}

\section{Introduction}
For the past decade, numerous deviations from Standard Model (SM) predictions have been observed in $B$ meson decays, particularly in $b \to s \ell^+ \ell^-$ transitions. Notably in the branching fractions and angular observables of $B$ decays to $M \mu^+ \mu^-$ ($M = K^{(*)}, \phi$) at low di-lepton momenta ($q^2$).

SM predictions for these observables are highly sensitive to non-perturbative contributions, which introduce significant uncertainties \cite{Mahmoudi:2024zna}. These contributions are divided into local and non-local types. Local contributions, represented by form factors, and non-local contributions, which are more difficult to quantify.

Most discrepancies manifest at low $q^2$, where computing hadronic form factors using QCD lattice \cite{Parrott:2022rgu} is particularly challenging. In this regime, QCD Light-Cone Sum Rules (LCSR) are employed~\cite{Khodjamirian:1997lay, Ball:2004rg, Ball:2004ye, Khodjamirian:2005ea, Khodjamirian:2006st, Bharucha:2015bzk, Khodjamirian:2017fxg, Gubernari:2018wyi,Cui:2022zwm,  Carvunis:2024koh}, although they entail systematic errors due to semi-local quark-hadron duality (QHD). We propose a strategy to bypass semi-global QHD, thereby exchanging unknown systematic errors for quantifiable and improvable errors from truncating the perturbative QCD and Light-Cone operator-product expansions (LCOPE).

\section{Light-Cone Sum Rules with \textit{\textbf{B}}-meson LCDAs}
To compute the form factors for $B \to M$ processes, we start with a $B$-meson to vacuum correlation function~\cite{Khodjamirian:2005ea, Khodjamirian:2006st, Gubernari:2018wyi,Cui:2022zwm, Monceaux:2023byy, Carvunis:2024koh}:
\begin{equation}\label{Correlation_function}
    \Pi^{\mu\nu}(q, k) = i \int d^4x e^{ik.x}\bra{0}T \left\{ J^\nu_{int}(x)J^\mu_{weak}(0) \right\} \ket{\bar{B}(p_B = q+k)},
\end{equation}
where $p_B$ is the $B$-meson 4-momentum and $q$ the momentum transfer. Here, $J^\mu_{weak}$ denotes a $b \to$ light transition current, $J^\nu_{int}$ is an interpolating current, and $x$ represents the time-space separation between these two currents.

By employing analyticity and a unitarity relation (via inserting a complete set of hadronic states), one can derive a relation for a given form factor $F$ of the form:
\begin{equation} \label{eq:Pi_p0}
    \Pi_{F}(q^2,k^2) = Y_F \frac{F(q^2)}{m_M^2 - k^2} + \int_{s_{cont}}^\infty \frac{\rho_F(q^2,s)}{s-k^2},
\end{equation}
where $s_{cont}$ is the threshold of the lowest continuum or excited state, $\rho_F$ is the spectral density, and $M$ is the lightest meson. The normalization factors $Y_F$ are expressed in terms of the masses and decay constants of the mesons involved. We refer to Table 1 of~\cite{Carvunis:2024koh} for their exact definitions. \\[1\baselineskip]
We then compute the $B$ to vacuum correlation function \eqref{Correlation_function} via a Light-Cone Operator Product Expansion (LCOPE)~\cite{Khodjamirian:1997lay} using $B$-meson LCDAs (Light-Cone Distribution Amplitudes) in growing twists. We work within the framework of HQET (Heavy Quark Effective Theory). The $B$-meson Fock state is expanded, retaining only the 2-particle and 3-particle contributions, and calculations are performed to leading order in QCD. This results in an expression we denote as $\Pi_F^{LCOPE}$.

\section{Quark-Hadron Duality} \label{sec:qhd}

With \textit{global} quark-hadron duality, we equate the scalar expressions $\Pi_F$ and $\Pi_F^{LCOPE}$. To extract the hadronic form factor $F$, the integral over the spectral density can be estimated using \textit{semi-global} quark-hadron duality. This approximation posits that for a sufficiently large negative $k^2$:
\begin{equation}\label{QH duality}
    \int_{s_{cont}}^{+ \infty}ds \frac{\rho_F(s)}{s-k^2} \approx \frac{1}{\pi}\int_{s_0}^{+ \infty}ds \frac{\text{Im}\Pi_F^{LCOPE}(s)}{s-k^2},
\end{equation}
where $s_0$, the quark-hadron duality threshold, is an effective parameter~\cite{Colangelo:2000dp}. The estimation of this parameter relies on the requirement that the value of the form factor is independent of $k^2$, forming a \textit{daughter sum rule}~\cite{Ball:2004rg, Ball:2004ye, Bharucha:2015bzk, Gubernari:2018wyi}. Alternatively, thresholds can be adopted from QCD sum rules~\cite{Khodjamirian:2006st, Wang:2015vgv, Gubernari:2018wyi}. \\[1\baselineskip]
This semi-global QHD approximation (Eq.~\eqref{QH duality}) introduces a theoretical uncertainty in predicting form factors. This is confirmed as the prediction of form factors depends critically on the strategy to determine the effective threshold $s_0$. We argue that this can yield particularly large errors for $B$-meson LCSRs.

We present a method to address this issue: we explore LCSRs in a regime where semi-global quark-hadron duality is unnecessary, thereby eliminating the associated systematic uncertainty. This approach is feasible, although it increases the dependence on radiative corrections and higher twists in the correlation function.

\section{Light-Cone Sum Rules without semi-global Quark-Hadron Duality}

Our objective is to eliminate the contribution from the spectral density integral \eqref{QH duality} to the correlation function $\Pi_F$ without employing semi-global QHD. Following~\cite{Shifman:1978bx}, we take the $p$-th derivative of Eq.~\eqref{eq:Pi_p0} with respect to $k^2$~\cite{Khodjamirian:2020btr}:
\begin{equation}
    \Pi_F^{(p)}(q^2,k^2) \equiv  \left( \frac{\partial}{\partial k^2} \right)^p \Pi_F(q^2,k^2) = \, p! \left( Y_F \frac{F(q^2)}{(m_M^2 - k^2)^{p+1}} + \int_{s_{cont}}^\infty \frac{\rho_F(s)}{(s-k^2)^{p+1}} \right).
\end{equation}
The form factor $F(q^2)$ can be rewritten as:
\begin{equation} 
    F(q^2) = \frac{(m_M^2 - k^2)^{p+1}}{p! \, Y_F} \Pi_F^{(p)}(q^2,k^2) - \int_{s_{cont}}^\infty \frac{\rho_F(s)}{Y_F} \left( \frac{m_M^2 - k^2}{s-k^2} \right)^{p+1}.
\end{equation}
Since $m_M^2 < s_{cont}$ and $k^2<0$:
\begin{equation}
    R_F \equiv \int_{s_{cont}}^\infty \frac{\rho_F(s)}{Y_F} \left( \frac{m_M^2 - k^2}{s-k^2} \right)^{p+1} \xrightarrow[p \to \infty]{} 0,
\end{equation}
the form factors can be expressed as:
\begin{equation}\label{eq:lim_FF}
    F(q^2) = \lim_{p\to\infty}  \frac{(m_M^2 - k^2)^{p+1}}{p! \, Y_F} \Pi_F^{(p)}(q^2,k^2).
\end{equation}
This expression is exact and does not rely on semi-global QHD. Similarly, an expression for the squared mass of the final meson can be derived as:
\begin{equation}\label{eq:lim_m2}
    m_M^2 = \lim_{p\to\infty}  \left[ \frac{p!}{(p-\ell)!}  \frac{\Pi_F^{(p-\ell)}}{\Pi_F^{(p)}} \right]^{1 / \ell} + k^2, \quad p>1 , \; p> \ell \geq 1 \,.
\end{equation}

For brevity, we define:
\begin{equation}
    \widetilde{\Pi}_F^{(p)}(q^2,k^2) \equiv \frac{(m_M^2 - k^2)^{p+1}}{p! \, Y_F} \Pi_F^{(p)}(q^2,k^2), \quad \widetilde{m}_M^2(p,\ell,k^2) \equiv \left[ \frac{p!}{(p-\ell)!}  \frac{\Pi_F^{(p-\ell)}}{\Pi_F^{(p)}} \right]^{1 / \ell} + k^2,
\end{equation}
such that
\begin{equation} 
    \widetilde{\Pi}^{(p)}_F(q^2,k^2) \xrightarrow[p \to \infty]{} F(q^2),  \quad \widetilde{m}_M^2(p,\ell,k^2) \xrightarrow[p \to \infty]{} m_M^2.
\end{equation}

These relations are exact, but require the correlation function $\Pi_F$ to all orders, while we only have access to the approximated expression expanded on the light-cone. Truncation errors grow as $p$ increases, affecting the accuracy. Our strategy is to quantify the $p$-dependent truncation error, ensuring $\Pi_F^{(p)} \approx \Pi_{F,LCOPE}^{(p)}$ within uncertainties. When $p$ becomes too large, the error diverges and the prediction loses predictive power. We aim to extract useful information from LCSRs without semi-global QHD if the truncation error remains manageable.

\begin{figure}[H]
    \centering
    \includegraphics[width=0.9\textwidth]{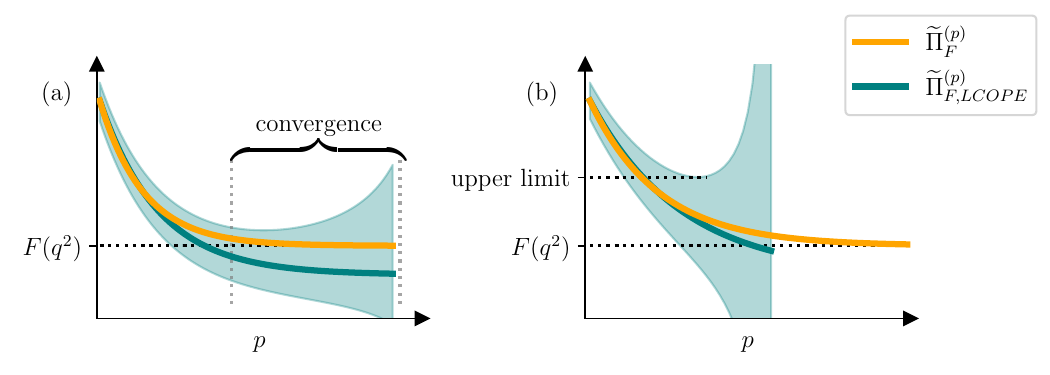}
    \caption{Schematic illustration of the strategy.}
    \label{fig:method_illustration}
\end{figure}
There are two eventual outcomes, as illustrated in Fig.\ref{fig:method_illustration}:
\begin{itemize}
    \item \underline{Convergence of the sum rule}: The spectral density integral $R_F$ becomes negligible before the truncation error diverges. The criteria for this regime are that $\widetilde{m}_M^2$ converges to $m_M^2$, and that we observe a weak dependence of both $\widetilde{\Pi}_{F,LCOPE}^{(p)}$ and $\widetilde{m}_M^2$ on $p$ and $k^2$.
    \item \underline{Upper limit}: The error diverges before reaching convergence. $R_F$ can be estimated with semi-global QHD, and if positive, this provides upper limits on form factors. Smaller SM predictions for $f_{+}^{B\to K}(q^2)$ could reduce the tension between experimental measurements and SM predictions of $\text{BR}(B^+ \to K^+ \mu^+ \mu^-)$.
\end{itemize}

This approach requires a conservative estimation of the truncation error for the multiple expansions, which include the HQET, the $B$-meson Fock state expansion, the radiative corrections, and the LCOPE.

\section{Numerical Results} \label{sec:numerics}
\begin{figure}[h]
    \centering
    \includegraphics[scale=0.27]{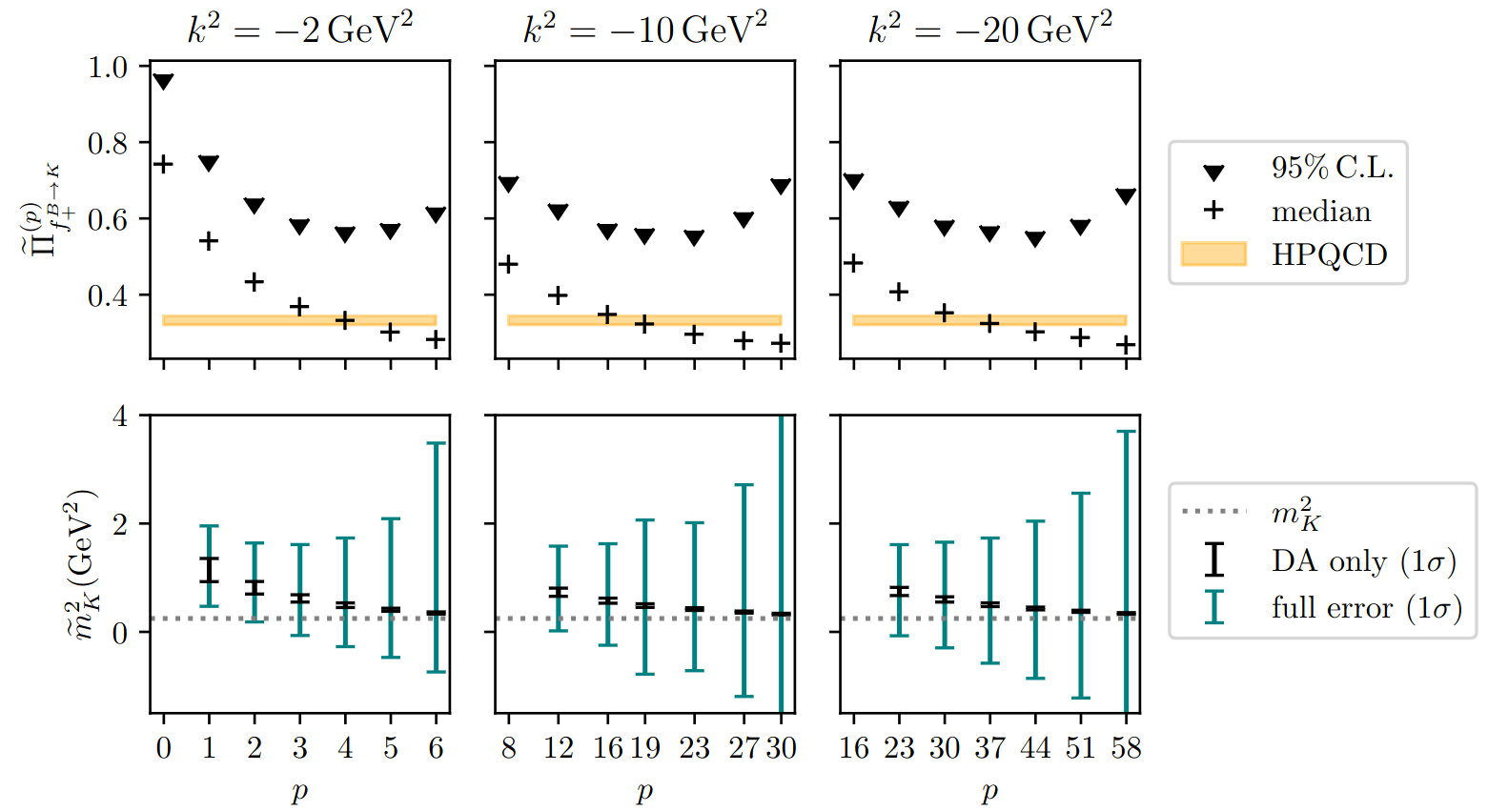}
    \caption{Numerical results for $k^2 = -2,-10,-20 \, \textrm{GeV}^2$ around the optimal value of $p$. \vspace{0.1cm} \\ 
        - $ 1^{\textrm{st}}$ row: $95^{\textrm{th}}$ percentile and median value of $\widetilde{\Pi}_{f_+}^{B \to K}$, $N=6000$. The orange bands represent the $1\sigma$ interval predicted by the HPQCD collaboration \cite{Parrott:2022rgu}.  \\
        - $2^{\textrm{nd}}$ row: blue (black) error bars: $1 \sigma$ intervals including all errors (parametric error only). 
        }
    \label{fig:fplusBtoK}
\end{figure}

We present results for $-k^2 = 2, 10, 20$ GeV$^2$ as representative values, and $q^2 = 0$. For each $-k^2$, we increase $p$ until the error in $\widetilde{\Pi}_{F,\, LCOPE}^{(p)}$ diverges, then sample points for a few values of $p$ before this divergence. This method is illustrated for the $f_+^{B \to K}$ form factor in Fig. \ref{fig:fplusBtoK}.

Our results for upper bounds and predictions are shown in Table \ref{tab:resultsBtoP}. Optimal $(k^2, p)$ values are chosen to obtain the most stringent $95\%$ confidence level upper limits for the form factors. We also provide the median value and the $1\sigma$ interval of $\widetilde{\Pi}_{F,\, LCOPE}^{(p)}$ at the optimal pair and estimate $R_F$ using semi-global QHD to test its sign and magnitude. The effective thresholds are taken from QCD sum rules.

The central values of $\widetilde{\Pi}_F^{(p)}$ indicate the form factor predictions assuming sum rule convergence. All predictions are compatible with existing literature but have large uncertainties. These uncertainties are mainly due to the error estimation from truncating perturbative QCD corrections and, to a lesser extent, the LCOPE.

\begin{table}[H]
    \centering
    \begin{tabular}{|c||c|c||c|c|c|c|}
    \hhline{-||--||----}
         Form Factor & $-k^2/p$ & $R_F(p,k^2)$ & \makecell{Upper Limit \\ @ $95\%$ C.L.} & $\widetilde{\Pi}_F^{(p)}$ $(1\sigma)$ & Literature & Ref.\\ \hhline{=::==::====}
          $f_+^{B\to K}$ & $10/19$ &  $0.03_{-0.01}^{+0.01}$ & $0.57$ & $0.32_{-0.12}^{+0.15}$ &  \makecell{$0.332(12)$ \\ $0.27(8)$ \\ $0.325(85)$ \\ $0.395(33)$ } & \makecell{\cite{Parrott:2022rgu} \\ \cite{Gubernari:2018wyi} \\ \cite{Cui:2022zwm} \\ \cite{Khodjamirian:2017fxg}  }\\ \hhline{-||--||----}
          $f_T^{B\to K}$ & $10/8$ & $0.04_{-0.06}^{+0.02}$ & $0.46$ & $0.34_{-0.07}^{+0.08}$ & \makecell{$0.332(21)$ \\ $0.25(7)$ \\ $0.381(27)$ \\ $0.381(97)$} & \makecell{\cite{Parrott:2022rgu} \\ \cite{Gubernari:2018wyi} \\ \cite{Khodjamirian:2017fxg} \\ \cite{Cui:2022zwm}} \\ \hhline{-||--||----}
        \end{tabular}
    \caption{Upper limits at the $95 \%$ confidence level and central value of $\widetilde{\Pi}_F^{(p)}$ for $B \to K$. We include the corresponding values of $-k^2$ (GeV$^2$) and $p$ as well as an estimate of $R_F(p,k^2)$ using semi-global quark-hadron duality.}
    \label{tab:resultsBtoP}
\end{table}

At this stage, we find that both criteria of convergence, the independence of the prediction with respect to $k^2$ and $p$, and the convergence of $\widetilde{m}_M^2$ to $m_M^2$, are respected within uncertainties. However, given the large uncertainties, this is not very useful for properly characterizing the convergence. These uncertainties are dominated by our QCD error estimation, and computing the radiative corrections should reduce this uncertainty considerably.

We make an interesting observation in the mass sum rule. As shown in Fig. \ref{fig:fplusBtoK}, the value of $\widetilde{m}^2_K$ including parametric errors only becomes remarkably close to the squared meson mass $m_K^2$ as $p$ increases, with very small parametric uncertainties. This is also true for $f_T^{B \to K}$. At this stage, this appears to be a numerical coincidence which has two possible explanations. Either the actual radiative corrections and higher-twist contributions are small, or more probably, they cancel out in the ratio \eqref{eq:lim_m2}. In any case, such a quick and accurate convergence seems to be a sign that $R_F \ll F(q^2)$ for a relatively large $-k^2/p$.

\section{Conclusion}

We propose a strategy to predict form factors using LCSRs without the semi-global QHD, thereby avoiding the determination of an effective threshold. Our approach involves conservatively evaluating the truncation error in the correlator and taking $-k^2/p$ to be as small as possible to suppress the spectral density integral $R_F$. We introduce two tests of the suppression of $R_F$ that do not rely on semi-global QHD: the $k^2$- and $p$-independence of the predicted form factors, and the convergence of the mass sum rule towards the physical value of the squared meson mass $\widetilde{m}_M^2 \to m_M^2$.

This approach significantly reduces the systematic uncertainty associated with the LCSR method. However, this improvement comes at the cost of a greater dependence on higher-order perturbative corrections and higher twists in the LCOPE. This trade-off is advantageous, as these corrections are calculable.

Currently, assessing the convergence of the sum rule in this regime is hampered by the relatively large truncation error affecting both $\widetilde{\Pi}_F^{(p)}$ and $\widetilde{m}_M^2$. However, we observe that without the modeled truncation error, $\widetilde{m}_M^2$ converges remarkably fast and accurately to the physical values of $m_M^2$ for $M = K$. This suggests that $R_F$ quickly converges to zero and that the errors due to the truncated contributions in pQCD and LCOPE cancel out in the mass sum rule ratio provided in Eq.~\eqref{eq:lim_m2}. In this case, computing the NLO QCD correction could demonstrate convergence in the region $-k^2/p \approx 0.5-0.3$ GeV$^2$.

\section*{Acknowledgments}

I am grateful to the organizers of Moriond QCD 2024 for the very interesting and useful conference. I also wish to thank A. Carvunis and F. Mahmoudi for their collaboration on this work.


\section*{References}

\begin{thebibliography}{99}
\bibitem{Mahmoudi:2024zna}
F.~Mahmoudi, and Y.~Monceaux, \Journal{Symmetry}{16}{1006}{2024}.

\bibitem{Parrott:2022rgu} 
W.~G.~Parrott, C.~Bouchard, and C.~T.~H.~Davies, \Journal{\PRD}{107}{014510}{2023}.

\bibitem{Khodjamirian:2005ea} 
A.~Khodjamirian, T.~Mannel, and N.~Offen, \Journal{\PLB}{620}{52--60}{2005}.

\bibitem{Khodjamirian:2006st} 
A.~Khodjamirian, T.~Mannel, and N.~Offen, \Journal{\PRD}{75}{054013}{2007}.

\bibitem{Gubernari:2018wyi} 
N.~Gubernari, A.~Kokulu, and D.~van Dyk, \Journal{\JHEP}{01}{150}{2019}.

\bibitem{Carvunis:2024koh} 
A.~Carvunis, F.~Mahmoudi, and Y.~Monceaux, \Journal{\PRD}{110}{114008}{2024}.

\bibitem{Khodjamirian:1997lay} 
A.~Khodjamirian, R.~Ruckl, S.~Weinzierl, C.~W.~Winhart, and O.~I.~Yakovlev, \Journal{"Adv. Ser. Direct. High Energy Phys."}{56}{2297--2321}{1997}.

\bibitem{Ball:2004rg} 
P.~Ball and R.~Zwicky, \Journal{\PRD}{71}{014029}{2005}.

\bibitem{Ball:2004ye} 
P.~Ball and R.~Zwicky, \Journal{\PRD}{71}{014015}{2005}.

\bibitem{Bharucha:2015bzk} 
A.~Bharucha, D.~M.~Straub, and R.~Zwicky, \Journal{\JHEP}{08}{098}{2016}.

\bibitem{Cui:2022zwm} 
B.-Y.~Cui, Y.-K.~Huang, Y.-L.~Shen, C.~Wang, and Y.-M.~Wang, \Journal{\JHEP}{03}{140}{2023}.

\bibitem{Khodjamirian:2017fxg} 
A.~Khodjamirian and A.~V.~Rusov, \Journal{\JHEP}{08}{112}{2017}.


\bibitem{Monceaux:2023byy} 
Y.~Monceaux, A.~Carvunis and F.~Mahmoudi, \Journal{PoS}{FPCP2023}{060}{2023}. 

\bibitem{Colangelo:2000dp}
P.~Colangelo and A.~Khodjamirian, {At the frontier of particle physics. Handbook of QCD. Vol. 1-3}{1495--1576}{2000}.

\bibitem{Wang:2015vgv} 
Y.-M.~Wang and Y.-L.~Shen, \Journal{\NPB}{898}{563--604}{2015}.

\bibitem{Shifman:1978bx} 
M.~A.~Shifman, A.~I.~Vainshtein, and V.~I.~Zakharov, \Journal{\NPB}{147}{385--447}{1979}.

\bibitem{Khodjamirian:2020btr}
A.~Khodjamirian, {Hadron Form Factors}: {From Basic Phenomenology to QCD Sum Rules} {2020}.


\end{thebibliography}
\end{document}